\pdfoutput=1

\documentclass[journal]{IEEEtran}

\usepackage{Preamble}

\newcommand{\Tensor}[1]{\mathbf{#1}}
\newcommand{\Rank}{\operatorname{rank}}
\newcommand{\Diagonal}{\operatorname{diag}}
\newcommand{\Row}{\operatorname{row}}
\newcommand{\Column}{\operatorname{col}}

\newcommand{\Set}[1]{\mathcal{#1}} 
\newcommand{\Cardinality}[1]{|#1|}

\newcommand{\Graph}[2]{(#1,#2)}
\newcommand{\Vertices}{\mathcal{V}}
\newcommand{\Edges}{\mathcal{E}}
\newcommand{\Incidence}{\Tensor{A}}

\newcommand{\Real}{\operatorname{Re}}

{
\theoremstyle{plain}
\newtheorem{Hypothesis}{Hypothesis}
\newtheorem{Lemma}{Lemma}
\newtheorem{Theorem}{Theorem}
\newtheorem{Corollary}{Corollary}
}

{
\theoremstyle{definition}
\newtheorem*{Proof}{Proof}
}

\begin{document}


\title
{%
	On the Properties of the\\Power Systems Nodal Admittance Matrix
}



\author{
	\IEEEauthorblockN{%
		Andreas~Martin~Kettner,~\IEEEmembership{Member,~IEEE},
		Mario~Paolone,~\IEEEmembership{Senior~Member,~IEEE}
	}
}

%
%


\markboth{}{}
%




\maketitle



\begin{abstract}
	This letter provides conditions determining the rank of the nodal admittance matrix,
	and arbitrary block partitions of it, for connected AC power networks with complex admittances.
	Furthermore, some implications of these properties concerning
	Kron Reduction and Hybrid Network Parameters are outlined.
\end{abstract}



%
\IEEEpeerreviewmaketitle

\section{Introduction}

%
%
%


\IEEEPARstart{T}{he} contributions of this letter are threefold.
First, the rank of the nodal admittance matrix of a generic AC power system with complex admittances
is determined in the absence and presence of shunt elements (Section \ref{Section:Rank}).
Second, it is shown that the diagonal blocks of the nodal admittance matrix given by
an arbitrary partition of the network's nodes always have full rank (Section \ref{Section:Invertibility}).
Third, some implications concerning the existence of the Kron Reduction and the Hybrid Network Parameters
are outlined (Section \ref{Section:Implications}).




\section{Foundations}
\label{Section:Foundation}


\subsection{Graph Theory}

Let $\Cardinality{\Set{S}}$ denote the \emph{cardinality} of a set $\Set{S}$.
A \emph{directed graph} with \emph{vertices} $\Set{V}=\{v_{1},\ldots,v_{\Cardinality{\Set{V}}}\}$
and \emph{edges} $\Set{E}=\{e_{1},\ldots,e_{\Cardinality{\Set{E}}}\}$ is denoted by $(\Set{V},\Set{E})$.
The connectivity of $(\Set{V},\Set{E})$ is defined by the \emph{incidence matrix} $\Tensor{A}_{(\Set{V},\Set{E})}$.
As known from graph theory \cite{Book:Desoer:1969:Circuit}%
\begin{Lemma}
	\label{Lemma:Incidence}
	If $(\Set{V},\Set{E})$ is connected, $\Rank(\Incidence_{\Graph{\Vertices}{\Edges}}) = \Cardinality{\Vertices}-1$.%
\end{Lemma}


\subsection{Circuit Theory}

Let $(\Set{V},\Set{E})$ define the topology of a power network.
Note that $\Set{V}=\Set{N}\cup\Set{G}$,
where $\Set{G}=\{v_{\Cardinality{\Set{V}}}\}$ is the reference (\emph{ground}) node,
and $\Set{N}=\{v_{1},\ldots,v_{\Cardinality{\Set{V}}-1}\}$ are generic \emph{nodes}.
All sources and nodal voltage phasors $V_{n}$ ($v_{n}\in\mathcal{N}$) are referenced to $\Set{G}$.
Let $\Tensor{V}_{\Set{N}}$ / $\Tensor{I}_{\Set{N}}$ be the vectors of nodal voltage / current phasors.
They are linked by the \emph{nodal admittance matrix}
$\Tensor{I}_{\Set{N}} = \Tensor{Y}_{\Set{N}}\Tensor{V}_{\Set{N}}$.
Connections between any pair of nodes in $\Set{N}$ are represented by
passive and reciprocal \emph{two-port equivalents}.
So $\Set{E}=\Set{L}\cup\Set{T}$, where $\Set{L}$ are the \emph{branches} and $\Set{T}$ are the \emph{shunts},
which correspond to the \emph{longitudinal} and \emph{transversal} electrical parameters of the lines, respectively.
Let $\Tensor{y}_{\Set{L}}$ and $\Tensor{y}_{\Set{T}}$ denote the associated admittances.
Then, $\Tensor{Y}_{\Set{N}}$ is given by \cite{Book:Desoer:1969:Circuit}%
\begin{equation}
	\Tensor{Y}_{\Set{N}}
	=	\Tensor{A}_{(\Set{N},\Set{L})}^{T}\Tensor{Y}_{\Set{L}}\Tensor{A}_{(\Set{N},\Set{L})} + \Tensor{Y}_{\Set{T}}
	\label{Equation:Admittance:Nodal}
\end{equation}
where $\Tensor{Y}_{\Set{T}}=\Diagonal(\Tensor{y}_{\Set{T}})$
with $\Rank(\Tensor{Y}_{\Set{T}})\leqslant\Cardinality{\Set{N}}$.
As for $\Tensor{Y}_{\Set{L}}$
\begin{Hypothesis}
	\label{Hypothesis:Admittance:Branch}
	The branches are not electromagnetically coupled and have nonzero admittance.
	Therefore, $\Tensor{Y}_{\Set{L}}=\Diagonal(\Tensor{y}_{\Set{L}})$
	with $\Rank(\Tensor{Y}_{\Set{L}})=\Cardinality{\Set{L}}$,
	where $\Tensor{y}_{\Set{L}}$ are the branch admittances.%
\end{Hypothesis}
\noindent
As known from circuit theory, it holds that \cite{Book:1990:Arrillaga:System}
\begin{Lemma}
	\label{Lemma:Admittance:Nodal}
	$\sum_{k}\Row_{k}(\Tensor{Y}_{\Set{N}})=\sum_{k}\Column_{k}(\Tensor{Y}_{\Set{N}}) =\Tensor{y}_{\Set{T}}$.
\end{Lemma}%


\subsection{Linear Algebra}

%
\begin{Lemma}
	\label{Lemma:Product:Symmetric}
	For any matrix $\Tensor{M}$, $\Rank(\Tensor{M}^{T}\Tensor{M})=\Rank(\Tensor{M})$.
\end{Lemma}
\begin{Lemma}
	\label{Lemma:Product}
	For square matrices $\Tensor{N}_{L}$,$\Tensor{N}_{R}$ with full rank and matching size, 
	$\Rank(\Tensor{N}_{L}\Tensor{M})=\Rank(\Tensor{M})=\Rank(\Tensor{M}\Tensor{N}_{R})$.
\end{Lemma}

\section{Rank}
\label{Section:Rank}

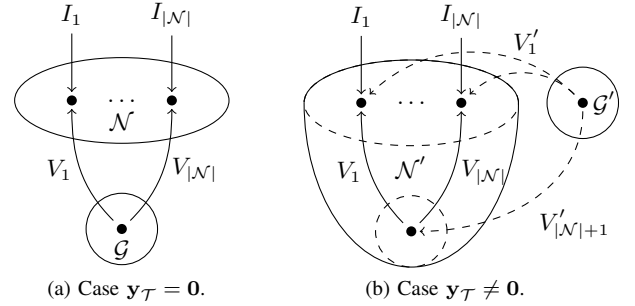
\begin{figure}[t]
	\centering
	\subfloat[Case $\Tensor{y}_{\Set{T}}=\Tensor{0}$.]
	{
		\centering
		\begin{tikzpicture}[scale=0.95]
	\small
		
	\def\xa{-0.7}
	\def\xb{0}
	\def\xc{+0.7}
	\def\ya{-1.8}
	\def\yb{0}
	\def\yc{+1.2}
	\def\dy{0.3}
	
	
	\node at (\xa,\yb) (N1) {};
	\node at (\xc,\yb) (N2) {};
	\node at (\xb,\yb) (N){$\ldots$};
	\node at (\xb,\ya) (G) {};
	
	\fill (N1) circle [radius=2pt];
	\fill (N2) circle [radius=2pt];
	\fill (G) circle [radius=2pt];
	
	\draw (G) [->,out=135,in=-90] to node[midway,left]{$V_{1}$} (N1);
	\draw (G) [->,out=45,in=-90] to node[midway,right]{$V_{\Cardinality{\Set{N}}}$} (N2);
	
	
	\node at (\xa,\yc) (I1) {$I_{1}$};
	\node at (\xc,\yc) (I2) {$I_{\Cardinality{\Set{N}}}$};
	
	\draw (I1) [->] to (N1);
	\draw (I2) [->] to (N2);
	
	
	\draw (N) ellipse (1.5 and 0.6);
	\draw (G) circle (0.5);
	
	\node at ($(N)-(0,\dy)$) {$\Set{N}$};
	\node at ($(G)-(0,\dy)$) {$\Set{G}$};
\end{tikzpicture}
		\label{Figure:Circuit:A}
	}
	\subfloat[Case $\Tensor{y}_{\Set{T}}\neq\Tensor{0}$.]
	{
		\centering
		\tikzstyle{block} = [draw, rectangle, minimum width = 0.75cm, minimum height = 0.75cm]
\tikzstyle{sum} = [draw, circle, minimum size=.5cm, node distance=1.75cm]
\tikzstyle{input} = [coordinate]
\tikzstyle{output} = [coordinate]

\tikzset{
    partial ellipse/.style args={#1:#2:#3}{
        insert path={+ (#1:#3) arc (#1:#2:#3)}
    }
}

\begin{tikzpicture}[scale=0.95]
	\small
		
	\def\xa{-0.7}
	\def\xb{0}
	\def\xc{+0.7}
	\def\xd{+2.4}
	\def\ya{-1.8}
	\def\yb{0}
	\def\yc{+1.2}
	\def\dy{0.3}
	
	
	\node at (\xa,\yb) (N1) {};
	\node at (\xc,\yb) (N2) {};
	\node at (\xb,\yb) (N){$\ldots$};
	\node at (\xb,\ya) (G1) {};
	\node at (\xd,\yb) (G2) {};
	
	\fill (N1) circle [radius=2pt];
	\fill (N2) circle [radius=2pt];
	\fill (G1) circle [radius=2pt];
	\fill (G2) circle [radius=2pt];
	
	\draw (G1) [->,out=134,in=-90] to node[midway,left]{$V_{1}$} (N1);
	\draw (G1) [->,out=45,in=-90] to node[midway,right]{$V_{\Cardinality{\Set{N}}}$} (N2);
	
	
	\draw[dashed] (G2) [->,out=135,in=45] to node [near start,above]{$V_{1}'$} (N1);
	\draw[dashed] (G2) [->,out=135,in=45] to (N2);
	\draw[dashed] (G2) [->,out=-90,in=0] to node[midway,below right]{$V_{\Cardinality{\Set{N}}+1}'$}  (G1);
	
	
	\node at (\xa,\yc) (I1) {$I_{1}$};
	\node at (\xc,\yc) (I2) {$I_{\Cardinality{\Set{N}}}$};
	
	\draw (I1) [->] to (N1);
	\draw (I2) [->] to (N2);
	
	
	\draw[dashed] (N) ellipse (1.5 and 0.6);
	\draw[dashed] (G) circle (0.5);
	\draw (G2) circle (0.5);
	\draw (N) [partial ellipse=0:180:1.5 and 0.6];
	\draw (N) [partial ellipse=180:360:1.5 and 2.3];
	
	\node at ($(\xb,\ya/2+\yb/2)$) {$\Set{N}'$};
	\node at ($(G2)+(\dy,0)$) {$\Set{G'}$};
\end{tikzpicture}
		\label{Figure:Circuit:B}
	}
	\caption{Proof of Theorem~\ref{Theorem:Rank} (surfaces show connected graphs).}
	\label{Figure:Circuit}
\end{figure}

\begin{Theorem}
	\label{Theorem:Rank}	
	If $(\Set{N},\Set{L})$ is connected and Hypothesis~\ref{Hypothesis:Admittance:Branch} holds
	\begin{equation}
		\Rank(\Tensor{Y}_{\Set{N}})
		=	\left\{
			\begin{array}{ll}
				\Cardinality{\Set{N}}-1	&	\text{if}~\Tensor{y}_{\Set{T}}=\Tensor{0}\\
				\Cardinality{\Set{N}}	& 	\text{otherwise}
			\end{array}
			\right.
	\end{equation}
	In other words, $\Tensor{Y}_{\Set{N}}$ has full rank if there is at least one shunt.
\end{Theorem}

The above property appears in the literature
(e.g. \cite{Conference:Krause:2011:LF:Probabilistic,Journal:Doerfler:2013:Graph:Kron}),
but an unobjectionable proof is not provided.
For instance, \cite{Journal:Doerfler:2013:Graph:Kron} relies on $\Tensor{Y}_{\Set{N}}$ being diagonally dominant,
which does not hold in general for complex admittances.
Therefore, a proof which works for the general case is presented in the following.
\begin{Proof}[$\Tensor{y}_{\Set{T}}=\Tensor{0}$, see Fig.~\ref{Figure:Circuit:A}]
	The claim follows from \eqref{Equation:Admittance:Nodal} using Lemma~\ref{Lemma:Product:Symmetric}, 
	\ref{Lemma:Product}, and \ref{Lemma:Incidence} ($\Tensor{Y}_{\Set{L}}=\Tensor{B}^{T}\Tensor{B}$,
	$\Tensor{M}=\Tensor{B}\Tensor{A}_{(\Set{N},\Set{L})}$).%
	\qed	
\end{Proof}

\begin{Proof}[$\Tensor{y}_{\Set{T}}\neq\Tensor{0}$, see Fig.~\ref{Figure:Circuit:B}]
	Let $\Set{G}'$ be a virtual ground.
	Define $\Set{N}'=\Set{N}\cup\Set{G}$, $\Set{L}'=\Set{L}\cup\Set{T}$, and $\Set{T}'=\emptyset$.
	Let $\Set{V}'=\Set{N}'\cup\Set{G}'$ and $\Set{E}'=\Set{L}'\cup\Set{T}'$
	form the graph $(\Set{V}',\Set{E}')$.
	Redefine the voltages
	\begin{equation}
		\renewcommand{\arraystretch}{1.3}
		\Tensor{V}_{\Set{N}'}'
		=
		\left[
		\begin{array}{c}
			\Tensor{V}'_{\Set{N}}	\\
			\hline
			V'_{\Cardinality{\Set{N}}+1}
		\end{array}
		\right]
		=	\left[
			\begin{array}{c|c}
				\Tensor{I}	&	\Tensor{1}	\\
				\hline
				\Tensor{0}	&	1
			\end{array}
			\right]
			\left[
			\begin{array}{c}
				\Tensor{V}_{\Set{N}}	\\
				\hline
				V'_{\Cardinality{\Set{N}}+1}
			\end{array}
			\right]
		\label{Equation:Voltage:Block}
	\end{equation}
	Note that the transformation matrix has full rank.
	Using $\Tensor{V}_{\Set{N}'}'$, the circuit equations may be written as follows
	\begin{equation}
		\renewcommand{\arraystretch}{1.3}		
		\left[
		\begin{array}{c}
			\Tensor{I}_{\Set{N}}	\\
			\hline
			\Tensor{I}_{\Cardinality{\Set{N}}+1}
		\end{array}
		\right]
		=
		\left[
		\begin{array}{r|r}
			\Tensor{Y}_{\Set{N}}		&	-\Tensor{y}_{\Set{T}}	\\
			\hline
			-\Tensor{y}_{\Set{T}}^{T}	&	Y_{\Set{T}}
		\end{array}
		\right]
		\left[
		\begin{array}{c}
			\Tensor{V}'_{\Set{N}}	\\
			\hline
			V'_{\Cardinality{\Set{N}}+1}
		\end{array}
		\right]
		\label{Equation:System:Block:1}
	\end{equation}
	where $Y_{\Set{T}}=\sum_{k}(\Tensor{y}_{\Set{T}})_{k}$.
	Eliminate $I_{\Cardinality{\Set{N}}+1}=-\sum_{k}(\Tensor{I}_{\Set{N}})_{k}$ (Kirchhoff's Law),
	substitute \eqref{Equation:Voltage:Block}, and use Lemma~\ref{Lemma:Admittance:Nodal}
	to obtain%
	\begin{equation}
		\renewcommand{\arraystretch}{1.2}		
		\left[
		\begin{array}{c}
			\Tensor{I}_{\Set{N}}	\\
			\hline
			0
		\end{array}
		\right]
		=
		\left[
		\begin{array}{c|c}
			\Tensor{Y}_{\Set{N}}	&	\Tensor{0}\\
			\hline
			\Tensor{0}					&	0
		\end{array}
		\right]
		\left[
		\begin{array}{c}
			\Tensor{V}_{\Set{N}}	\\
			\hline
			V'_{\Cardinality{\Set{N}}+1}
		\end{array}
		\right]
		\label{Equation:System:Block:2}
	\end{equation}
	According to Theorem~\ref{Theorem:Rank}, the matrix in \eqref{Equation:System:Block:1}
	has rank $\Cardinality{\Set{N}}$, since $(\Set{N}',\Set{L}')$ is connected and
	$\Tensor{y}_{\Set{T}'}=\Tensor{0}$.
	The elimination of $I_{\Cardinality{\Set{N}}+1}$ solely requires elementary row operations,
	which preserve the rank.
	Due to Lemma~\ref{Lemma:Product}, the change of coordinates \eqref{Equation:Voltage:Block}
	preserves the rank, too.
	Thus, the matrices in \eqref{Equation:System:Block:1} and \eqref{Equation:System:Block:2}
	have the same rank $|\Set{N}|$, which proves the claim $\Rank(\Tensor{Y}_{\Set{N}})=\Cardinality{\Set{N}}$.%
	\qed
\end{Proof}

\section{Block Rank}
\label{Section:Invertibility}

\begin{figure}[t]
	\centering
	\tikzset{
    partial ellipse/.style args={#1:#2:#3}{
        insert path={+ (#1:#3) arc (#1:#2:#3)}
    }
}

%
%
%
%
%
%
%

\begin{tikzpicture}[scale=0.95]
	\small
		
	
	\def\xa{-1.6}
	\def\xb{+1.6}
	\def\y{0}
	\def\rp{1.5}	
	\def\rc{0.5}	
	
	
	\node at (\xa,\y) (NP) {$\Set{N}_{p}$};
	\node at (\xb,\y) (NQ) {$\Set{N}_{q}$};
	
	\draw[dotted] (NP) circle (\rp);
	\draw[dotted] (NQ) circle (\rp);
	
	
	\node at ($(NP)+(0:0.9)$) (NPC) {$\Set{N}_{p,c}$};	
	\node at ($(NQ)-(0:0.9)$) (NQC) {$\Set{G}$};
	\draw[dashed] (NQC) circle (\rc);
	
	\foreach \a in {60,180,300} {
		\draw[dotted] (NP) to ($(NP)+(\a:\rp)$);
	}
	
	\foreach \a in {0,120,240} {
		\draw[dashed] ($(NP)+(\a:0.9)$) circle (\rc);
	}
	
	
	\draw ($(\xa/2+\xb/2,\y)$) ellipse (1.3 and 0.8);
\end{tikzpicture}
	\caption{Proof of Theorem~\ref{Theorem:Rank:Block} (surfaces show connected graphs).}
	\label{Figure:Component}
\end{figure}
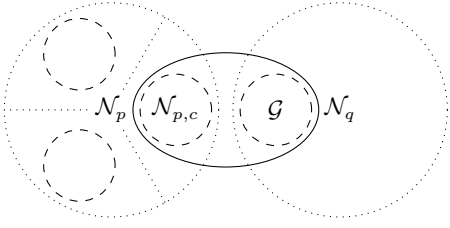

Let $\{\Set{N}_{p}\}$ ($p\in\Set{P}$) be a partition of $\Set{N}$
into $\Cardinality{\Set{P}}\geqslant2$ subsets.
Bring $\Tensor{Y}_{\Set{N}}$ into block form by reordering its rows and columns
and let $\Tensor{Y}_{\Set{N},ij}$ denote the block of $\Tensor{Y}_{\Set{N}}$ relating
$\Tensor{I}_{\Set{N},i}$ and $\Tensor{V}_{\Set{N},j}$%
\begin{Theorem}
	\label{Theorem:Rank:Block}
	If $(\Set{N},\Set{L})$ is connected, Hypothesis~\ref{Hypothesis:Admittance:Branch} holds,
	and all branches have $\Real((\Tensor{y}_{\Set{L}})_{l})>0$ ($\forall e_{l}\in\Set{L}$),
	then all diagonal blocks $\Tensor{Y}_{\Set{N},pp}$ ($p\in\Set{P}$) of $\Tensor{Y}_{\Set{N}}$ have full rank.
\end{Theorem}
\begin{Proof}
	$\Tensor{Y}_{\Set{N},pp}$ links $\Tensor{I}_{\Set{N},p}$ and $\Tensor{V}_{\Set{N},p}$
	for  $\Set{N}_{q}$ ($q\in\Set{P}$, $q\neq p$) grounded.
	i.e. branches between $\Set{N}_{p}$ and $\Set{N}_{q}$ act as shunts.
	Let $(\Set{N}_{p},\Set{L}_{p})$ be the part of $(\Set{N},\Set{L})$ associated with $\Set{N}_{p}$.
	%
	As $(\Set{N},\Set{L})$ is connected, $(\Set{N}_{p},\Set{L}_{p})$ consists of mutually disconnected
	\emph{components} $(\Set{N}_{p,c},\Set{L}_{p,c})$ ($c\in\Set{C}$),
	so $\Tensor{Y}_{\Set{N},pp}=\Diagonal(\{\Tensor{Y}_{\Set{N},pp,cc}\})$,
	and $\exists e_{l}\in\Set{L}$ with one end at $v_{n}\in\Set{N}_{p,c}$,
	acting as shunt $\forall c\in\Set{C}$ (Fig.~\ref{Figure:Component}).
	Since $\Real((\Tensor{y}_{\Set{T}})_{n})\geqslant0$ (passivity) and $\Real((\Tensor{y}_{\Set{L}})_{l})>0$,
	the total shunt admittance at $v_{n}$ is nonzero.
	By Theorem~\ref{Theorem:Rank}, all $\Tensor{Y}_{\Set{N},pp,cc}$ have full rank,
	so $\Tensor{Y}_{\Set{N},pp}$ has full rank, too.%
	\qed
\end{Proof}

\section{Implications}
\label{Section:Implications}


\subsection{Kron Reduction}
\label{Section:C:Kron}

\begin{Corollary}
	\label{Corollary:Kron}
	Suppose that the assumptions of Theorem~\ref{Theorem:Rank:Block} hold.
	Let $\Set{N}_{t}$ ($t\in\Set{P}$) be a set of zero injection nodes, i.e. $\Tensor{I}_{\Set{N},t}=\Tensor{0}$.
	Then, the $\Tensor{V}_{\Set{N},s}$ ($s\in\Set{P}$, $s\neq t$) uniquely define $\Tensor{V}_{\Set{N},t}$,
	so that $\Tensor{I}_{\Set{N}}=\Tensor{Y}_{\Set{N}}\Tensor{V}_{\Set{N}}$ may be reduced without loss of information. 
\end{Corollary}
\noindent
This technique, also known as \emph{Kron Reduction} \cite{Book:Kron:1959:Circuits},
allows to reduce the number of independent variables in the model of an electrical network.
Thereby, computationally heavy tasks like Power Flow computations
or State Estimation \cite{Book:Abur:2004:Estimation} may be considerably accelerated.
Although this technique is widely used in the field, researchers hardly ever verify whether the reduction is indeed feasible.
For instance, the inventor $\cite{Book:Kron:1959:Circuits}$ does not consider this issue at all,
and \cite{Journal:Doerfler:2013:Graph:Kron} only examines simple cases (purely resistive / inductive connections).
In this regard, Corollary~\ref{Corollary:Kron} ensures that Kron Reduction can be performed.
\begin{Proof}
	Expand block row $t$ of $\Tensor{I}_{\Set{N}}=\Tensor{Y}_{\Set{N}}\Tensor{V}_{\Set{N}}$
	\begin{equation}
		\Tensor{I}_{\Set{N},t}
		=		\Tensor{Y}_{\Set{N},tt}\Tensor{V}_{\Set{N},t}
			+	\sum\nolimits_{k\neq t}\Tensor{Y}_{\Set{N},tk}\Tensor{V}_{\Set{N},k} = \Tensor{0}
		\label{Equation:Current:Tie}
	\end{equation}
	Theorem~\ref{Theorem:Rank:Block} states that $\Tensor{Y}_{\Set{N},tt}$ has full rank,
	so \eqref{Equation:Current:Tie} may be solved for
	$\Tensor{V}_{\Set{N},t}=-\Tensor{Y}_{\Set{N},tt}^{-1}\sum_{k\neq t}\Tensor{Y}_{\Set{N},tk}\Tensor{V}_{\Set{N},k}$.
	Substitute $\Tensor{V}_{\Set{N},t}$ into $\Tensor{I}_{\Set{N},s}=\sum_{k}\Tensor{Y}_{\Set{N},sk}\Tensor{V}_{\Set{N},k}$
	($s\in\Set{P}$, $s\neq t$) to obtain
	\begin{align}
		\Tensor{I}_{\Set{N},s}
		&=	\sum\nolimits_{k\neq t}\widehat{\Tensor{Y}}_{\Set{N},sk}\Tensor{V}_{\Set{N},k}
		\\
		\widehat{\Tensor{Y}}_{\Set{N},sk}
		&=	\Tensor{Y}_{\Set{N},sk}
				-\Tensor{Y}_{\Set{N},st}\Tensor{Y}_{\Set{N},tt}^{-1}\Tensor{Y}_{\Set{N},tk}
	\end{align}
	which defines the reduced nodal admittance matrix $\widehat{\Tensor{Y}}_{\Set{N}}$.
	\qed
\end{Proof}


\subsection{Hybrid Network Parameters}
\label{Section:C:Hybrid}

\begin{Corollary}
	\label{Corollary:Hybrid}
	If the assumptions of Theorem~\ref{Theorem:Rank:Block} hold, then one can solve
	block row $p$ of $\Tensor{I}_{\Set{N}}=\Tensor{Y}_{\Set{N}}\Tensor{V}_{\Set{N}}$
	for $\Tensor{V}_{\Set{N},p}$ $\forall p\in\Set{P}$.
	Thus, there exists a hybrid network parameter matrix $\Tensor{H}$.%
\end{Corollary}

\noindent
The existence of hybrid network parameters has for instance been investigated in
\cite{Journal:Zuidweg:1965:Multiport:Hybrid,Journal:Anderson:1966:Multiport:Hybrid},
but the obtained criteria are not straightforward to apply to power systems.
One application lies in \emph{Voltage Stability Assessment},
namely some \emph{Voltage Stability Indices} \cite{Journal:Kessel:1986:Stability:Index,Journal:Wang:2013:Stability:Index}.
More precisely, the hybrid network parameters are used to establish the link between
the buses whose voltage is regulated and those where power is absorbed or injected.
In this regard, Corollary~\ref{Corollary:Hybrid} guarantees the existence of the required $\Tensor{H}$ matrix.

\begin{Proof}
	\label{Proof:Kron}
	Theorem~\ref{Theorem:Rank:Block} guarantees that $\Tensor{Y}_{\Set{N},pp}$ has full rank,
	so one can solve $\Tensor{I}_{\Set{N},p}=\sum_{k}\Tensor{Y}_{\Set{N},pk}\Tensor{V}_{\Set{N},k}$
	for $\Tensor{V}_{\Set{N},p}$, which yields
	\begin{align}
		\Tensor{V}_{\Set{N},p}
		&=	\Tensor{H}_{pp}\Tensor{I}_{\Set{N},p}+\sum\nolimits_{k\neq p}\Tensor{H}_{pk}\Tensor{V}_{\Set{N},k}
		\\
		\Tensor{H}_{pk}
		&=	\left\{
				\begin{array}{cc}
					\Tensor{Y}_{\Set{N},pp}^{-1}									&(k=p)\\
					-\Tensor{Y}_{\Set{N},pp}^{-1}\Tensor{Y}_{\Set{N},pk}	&(k\neq p)
				\end{array}
				\right.
	\end{align}
	Substitute $\Tensor{V}_{\Set{N},p}$ into
	$\Tensor{I}_{\Set{N},q}=\sum_{k}\Tensor{Y}_{\Set{N},qk}\Tensor{V}_{\Set{N},k}$ ($q\in\Set{P}$, $q\neq p$)
	\begin{align}
		\Tensor{I}_{\Set{N},q}
		&=	\Tensor{H}_{qp}\Tensor{I}_{\Set{N},p}+\sum\nolimits_{k\neq p}\Tensor{H}_{qk}\Tensor{V}_{\Set{N},k}
		\\
		\Tensor{H}_{qk}
		&=	\left\{
				\begin{array}{cc}
					\Tensor{Y}_{\Set{N},qp}\Tensor{Y}_{\Set{N},pp}^{-1}																		&(k=p)\\
					\Tensor{Y}_{\Set{N},qk}-\Tensor{Y}_{\Set{N},qp}\Tensor{Y}_{\Set{N},pp}^{-1}\Tensor{Y}_{\Set{N},pk}	&(k\neq p)
				\end{array}
				\right.
	\end{align}
	Thus, there obviously exists a matrix $\Tensor{H}$ as claimed.%
	\qed
\end{Proof}



\ifCLASSOPTIONcaptionsoff
	\newpage
\fi



%

\bibliographystyle{IEEEtran}
\bibliography{Bibliography}

\clearpage

\end{document}